\documentclass[11pt]{article}

\usepackage{physics}
\usepackage{braket}
\usepackage{mathtools}
\usepackage[margin=1in]{geometry}
\usepackage{amsmath,amssymb,amsfonts,setspace,url,mathrsfs}
\usepackage{bm}
\usepackage{algpseudocode}
\usepackage{enumitem}
\usepackage{latexsym}
\usepackage{amsthm}
\usepackage{subfigure}
\usepackage{float}
\usepackage{stmaryrd}
\usepackage{fancybox}
\usepackage{bigstrut, array, multirow}
\usepackage{here}
\usepackage{color}
\usepackage{url}
\usepackage[unicode, psdextra]{hyperref}
\usepackage{mathpazo}
\usepackage{thmtools}
\usepackage{thm-restate}
\usepackage{aliascnt}
\usepackage[capitalise]{cleveref}
\usepackage[normalem]{ulem}

\usepackage{xcolor}
\usepackage{tikz}
\usetikzlibrary{arrows.meta,calc,positioning,decorations.pathreplacing}
\definecolor{proofInk}{HTML}{454240}   
\definecolor{proofBlue}{HTML}{6890B9}  
\definecolor{proofGreen}{HTML}{87C9A0} 
\definecolor{proofRose}{HTML}{B97780}  
\definecolor{proofGold}{HTML}{E0A55C}  
\tikzset{
  proof picture/.style={font=\small,text=proofInk,line cap=round,line join=round},
  proof vertex/.style={circle,draw=proofInk!65,fill=white,minimum size=5.4mm,
    inner sep=1pt,font=\small},
  proof leader/.style={proof vertex,draw=proofGreen,fill=proofGreen!18,
    line width=1pt,minimum size=7.6mm},
  proof dot/.style={circle,fill=proofBlue,inner sep=1.5pt},
  proof region/.style={rounded corners=7pt,draw=proofBlue!60,fill=proofBlue!10,
    line width=.75pt},
  proof child/.style={rounded corners=7pt,draw=proofGreen!65,fill=proofGreen!12,
    line width=.75pt},
  proof lower/.style={rounded corners=7pt,draw=proofGold!65,fill=proofGold!11,
    line width=.75pt},
  proof edge/.style={draw=proofBlue,line width=1.1pt},
  proof path/.style={proof edge,line width=1.2pt,
    dash pattern=on 1pt off 3pt,line cap=round},
  proof choice/.style={draw=proofRose,line width=1.2pt,-{Latex[length=1.7mm]}},
  proof bridge/.style={draw=proofGreen,line width=1.35pt,-{Latex[length=1.7mm]}},
  proof crossing/.style={draw=proofGold,line width=1.1pt,dashed},
  proof removed/.style={draw=proofRose,line width=1.05pt,dashed},
  proof note/.style={font=\footnotesize,align=center,text=proofInk},
  proof heading/.style={font=\small\bfseries,align=center},
  proof tag/.style={font=\footnotesize,fill=none,inner sep=2pt},
  proof flow/.style={draw=proofInk!50,line width=.8pt,-{Latex[length=2mm]}}
}

\hypersetup{hidelinks}

\makeatletter
\renewcommand\thmt@autorefsetup{%
  \@xa\def\csname\thmt@envname autorefname\@xa\endcsname\@xa{\thmt@thmname}%
}
\makeatother

\newtheorem{theorem}{Theorem}
\newaliascnt{lemma}{theorem}
\newtheorem{lemma}[lemma]{Lemma}
\aliascntresetthe{lemma}
\newaliascnt{corollary}{theorem}
\newtheorem{corollary}[corollary]{Corollary}
\aliascntresetthe{corollary}
\crefname{lemma}{Lemma}{Lemmas}
\Crefname{lemma}{Lemma}{Lemmas}
\crefname{corollary}{Corollary}{Corollaries}
\Crefname{corollary}{Corollary}{Corollaries}

\DeclareMathOperator*{\EE}{\mathbb{E}}
\newcommand{\CC}{\mathcal{CC}}
\newcommand{\cc}{\mathrm{cc}}
\newcommand{\Crossing}{X}
\newcommand{\crossing}{x}
\newcommand{\Br}{\operatorname{Br}}
\newcommand{\enlc}{L}
\newcommand{\OmittedCount}{\eta}

\begin{document}

\hypersetup{pageanchor=false}
\begin{titlepage}
	
	\title{Two choices suffice: $k$-out sampling leaves $O(n/k)$ intercomponent edges for $k\geq 2$}

\author{Yanyu Chen\thanks{National University of Singapore, Singapore. Email: {\tt yanyu.chen@u.nus.edu}.}
\and Yu Chen\thanks{National University of Singapore, Singapore. Email: {\tt yu.chen@nus.edu.sg}.}} 

	\maketitle
	
	\thispagestyle{empty}
    
\begin{abstract}
    Given a simple undirected graph on $n$ vertices, random $k$-out sampling forms a spanning subgraph by having each vertex independently choose $k$ incident edges uniformly with replacement. We prove that, for every integer $k\geq2$, the expected number of original edges joining distinct sampled components is $O(n/k)$.

    This resolves the conjecture of Holm, King, Thorup, Zamir, and Zwick in its full range, improving their logarithmic sampling requirement to the optimal threshold of two choices per vertex. The bound fails for $k=1$, and the dependence $O(n/k)$ is tight up to constant factors. Our result holds with replacement and consequently also without replacement.
    In particular, this implies that contracting just two independent local choices per vertex suffices to reduce an arbitrary simple graph to a quotient with $O(n)$ edges in expectation, which is crucial in reducing the instance size for many contraction-based algorithms.

    Our proof uses switching arguments to relate intercomponent edges to bridges and a degree-weighted component potential, whose expectation we also bound with a switching argument. 
\end{abstract}

\end{titlepage}
\hypersetup{pageanchor=true}

\tableofcontents

\newpage

\section{Introduction}

Random edge sampling is a bsimple tool for sparsifying graphs while preserving useful structural information. We study sampling based on local choices at each vertex and ask how many original edges join distinct components of the sample. 
This property concerns connectivity of the original graph as contracting each sampled component leaves precisely these \emph{intercomponent edges} and adding them to a spanning forest of the sample preserves the original connected components. Bounds on the number of intercomponent edges are therefore useful for contraction-based minimum-cut algorithms~\cite{ghaffari2020faster}, spanning-forest communication protocols~\cite{holm2019random}, and spanning adjacency oracles~\cite{bodwin2023spanning}.

Let $G=(V_G, E_G)$ be a simple undirected graph on $n$ vertices. 
Holm, King, Thorup, Zamir, and Zwick~\cite{holm2019random} studied this question in the without-replacement $k$-out model, where each vertex $v$ independently chooses a uniformly random subset of $\min\{k,\deg_G(v)\}$ incident edges and the sampled subgraph is the union of these choices.
They proved that the expected number of intercomponent edges is $O(n/k)$ for $k\geq \Omega(\log n)$ and conjectured that the same bound should hold for all $k\geq 2$.

We resolve this conjecture in full, for every integer $k\geq 2$, in the stronger \emph{with-replacement} model.
Here each vertex independently samples $k$ incident edges uniformly with replacement, and $H_G^k$ is the union of these choices.
Writing $\crossing_G(H)$ for the number of intercomponent edges of a spanning subgraph $H$, we prove the following result.

\begin{theorem}
\label{thm: geq2-out crossing}
    Let $G$ be a simple graph on $n$ vertices, and let $H_G^k$ be its random with-replacement $k$-out subgraph. For every integer $k\geq 2$,
\[
    \EE\left[\crossing_G(H_G^k)\right]
    \leq O(n/k).
\]
\end{theorem}

The bound also holds without replacement as under a natural coupling, the without-replacement subgraph contains $H_G^k$ and has no more intercomponent edges (see \Cref{wor-vs-wr}). 
In this work, we use the with-replacement model unless stated otherwise.

The result is optimal in both its sampling rate $k$ and its intercomponent edge count. 
The bound fails for $k=1$, while examples of Holm, King, Thorup, Zamir, and Zwick~\cite{holm2019random} have $\Omega(n/k)$ expected intercomponent edges for $2\leq k\leq n/3$, even without replacement.

\subsection{Technical Contribution and Significance}

A key obstacle to extending the analysis of Holm, King, Thorup, Zamir, and Zwick~\cite{holm2019random} to constant $k$ is their reduction to the expected $k$-out model, where each vertex $v$ independently includes each incident edge with probability $\frac{k}{\max\{k,\deg(v)\}}$.
For constant $k$, this model leaves each vertex of a complete graph isolated with constant probability, yielding $\Omega(n^2)$ expected intercomponent edges instead of $O(n)$.

Our analysis bypasses this intermediate model and analyzes the $k$-out model directly using the \emph{switching technique}. 
The switching method is a double-counting technique for estimating the relative sizes of classes of combinatorial objects. We define transformations called switchings between the classes and compare the numbers of forward and inverse operations, thereby obtaining counting and probability bounds. 
In graph applications, a switching typically makes local modifications to the graphs, replacing a small set of edges with another. This is not necessary, but it often makes counting and reasoning about the number of forward and inverse switchings simpler. 
We use one switching argument to relate intercomponent edges to a component potential, and another to bound the expectation of this potential.

The case $k\geq3$ is outlined in \Cref{sec:technical-overview} and proved in \Cref{sec: general case k3}. The two-out case in \Cref{thm:two-out} follows the same two steps but requires more careful treatment.

\subsection{Implications and Applications} 

While the distributed sketching application of Holm, King, Thorup, Zamir, and Zwick~\cite{holm2019random} is already supported by their bound for $k\geq \Omega(\log n)$, extending the bound to constant $k$ enables further algorithmic applications.

In particular, our theorem provides the simplification anticipated by Ghaffari, Nowicki, and Thorup ~\cite{ghaffari2020faster} in their contraction-based algorithms for exact minimum cut in simple unweighted graphs. 
Their random $2$-out contraction reduces the graph to $O(n/\delta)$ vertices, where $\delta$ is the minimum degree, while preserving any fixed nontrivial minimum cut with constant probability. 
However, they have no bound on the number of edges that remain, and so they apply an additional sparsification step to reduce the edge count to $O(n)$. 
Our theorem supplies this edge bound directly for their random $2$-out contraction with replacement, removing the need for the additional edge sparsification step.
Combined with their vertex bound, it yields, with constant probability, a contracted instance with $O(n/\delta)$ vertices and $O(n)$ edges that preserves a fixed non-singleton minimum cut. 
Keeping $k$ constant is essential here, since it is what gives the constant cut-survival probability.

The same contraction principle improves exact minimum cut in the cut-query model, where each query returns the number of edges crossing a specified vertex subset. 
Kenneth-Mordoch and Krauthgamer ~\cite{kennethmordoch_et_al:LIPIcs.ESA.2025.100} obtained a two-round algorithm using $\widetilde O(n^{4/3})$ queries by constructing a contraction in the first round and recovering the contracted graph in the second. 
The recovery cost is nearly linear in the number of remaining vertices and edges. 
Our expected edge bound lets these sampling and recovery primitives implement random $2$-out contraction and reconstruct the contracted graph using $\widetilde O(n)$ queries, with the cut-survival guarantee above.

\subsection{Related Work} 


Early work focused on $k$-out sampling in the complete graph. Fenner and Frieze ~\cite{fenner1982connectivity} established $k$-vertex and $k$-edge connectivity for $k\geq 2$, Frieze ~\cite{frieze1986maximum} proved the existence of perfect matchings for even-order graphs when $k\geq 2$, and Bohman and Frieze ~\cite{bohman2009hamilton} proved Hamiltonicity for $k\geq 3$, all with probability tending to $1$.
These results can be found in the book by Frieze and Karo\'{n}ski ~\cite{frieze2015introduction}.

Frieze and Johansson ~\cite{frieze2017random} extended connectivity and Hamiltonicity results to graphs $G$ whose minimum degree is $(1/2+\epsilon)n$ for some $\epsilon>0$. They showed that $H^k_G$ is $k$-connected w.h.p. for $k = O(1)$ and is Hamiltonian for sufficiently large $k$. However, they remarked that such results are not possible for graphs with smaller minimum degree.

\paragraph{Minimum cut.}
Beyond the cut-query application above, random $2$-out contraction has also been used in streaming. Assadi and Dudeja~\cite{assadi2021simple} gave a two-pass algorithm for exact global minimum cut using $O(n\log n)$ bits of space, whose first pass builds a with-replacement $2$-out sample.

\paragraph{Spanning structures.}
The bound of Holm, King, Thorup, Zamir, and Zwick~\cite{holm2019random} also gives a one-way communication protocol for finding a spanning forest, using private randomness and $O(\sqrt{n}\log n)$ bits per vertex.
Bodwin and Fleischmann~\cite{bodwin2023spanning} also apply this bound to spanning adjacency oracles, which implicitly represent a sparse spanning subgraph and answer edge-membership queries. One construction uses $k$-out sampling. 

\paragraph{Beyond connectivity.}
Saneian and Behnezhad~\cite{saneian2024streaming} use one-sided random $3$-out sampling in bipartite graphs for streaming edge coloring. Each vertex on one side independently chooses three distinct uniform neighbors on the other. When the opposite side is larger by a sufficiently large constant factor, the sample contains a matching covering the sampling side with high probability.

\paragraph{Implementations.}
The ConnectIt framework of Dhulipala, Hong, and Shun~\cite{dhulipala2020connectit} and the GPU study of Hong, Dhulipala, and Shun~\cite{hong2020exploring} evaluate $k$-out sampling and its variants for parallel connectivity.

\paragraph{The switching method.}
As described in Wormald’s survey \cite{wormald1999modelssurvey}, McKay introduced switchings into the study of random graphs to bound subgraph probabilities in random graphs with prescribed degrees ~\cite{mckay1981subgraphs}. 
Since then, the switching argument has found many applications, including cycle counts and girth ~\cite{mckay2004short}, connectivity ~\cite{cooper2002random}, asymptotic estimates of spanning trees ~\cite {mckay1981spanning}, and asymptotic enumeration of graphs with prescribed degrees ~\cite{mckay1991asymptotic}. 
Switchings are also crucial in the analysis of exact uniform generation algorithms, beginning with the work of McKay and Wormald \cite{MW1990uniform} which was later improved by Arman, Gao, and Wormald \cite{AGW2019} also using switching. 


\section{Technical Overview}
\label{sec:technical-overview}

We prove the bound by relating intercomponent edges to a degree-sensitive potential of the sampled components, and then showing that a special augmentation of a $2$-out sample has expected potential $O(n)$. 

\subsection{From intercomponent edges to component potential}
We first describe a simple switching argument to demonstrate how it can be useful.
Fix $r<k$, and leave the first $r$ choices at every vertex unchanged.
For each intercomponent edge $e$ of $H^k$, choose its smaller-degree endpoint $v$ and replace any one of its remaining $k-r$ choices by $e$.
Each replacement makes $e$ a bridge of the new sample outside the unchanged $H^r$.

Each intercomponent edge therefore gives $k-r$ forward switchings.
Conversely, an image bridge at $v$ admits at most $\deg_G(v)$ inverse switchings, one for each possible previous choice.
Thus forward switchings count intercomponent edges multiplied by $k-r$, while inverse switchings are bounded by bridges weighted by the smaller degree of their endpoints.

These bridge weights are controlled by the component potential
\[
    \Phi_G(H^r)
    :=\sum_{C\in\CC(H^r)}\max_{v\in C}\deg_G(v).
\]
Indeed, after contracting the components of $H^r$, the bridges outside $H^r$ form a forest.
Rooting its trees and charging each bridge to its child component results in at most one charge per component.
Each charge is at most that component's maximum original degree.
Double-counting the switchings therefore gives
\[
    (k-r)\EE[\crossing_G(H^k)]
    \leq \EE[\Phi_G(H^r)].
\]
It remains to bound the expected potential of a suitable subgraph by $O(n)$.

\subsection{Bounding the component potential}
Fix a total order on the vertices by nondecreasing degree, breaking ties arbitrarily. 
Call the largest vertex of each component its \emph{leader}, and order the components by their leaders. 
The potential $\Phi_G(H)$ is the sum of the leaders' degrees. Their internal neighbors contribute at most $n$ because the components are disjoint, so it remains to control their crossing edges.
Consider each crossing edge to be \emph{owned} by the endpoint in the higher component.
We prove a bound on the crossing edges owned by leaders using a recurrence. 

\paragraph{A recurrence for edges at successive depths.}

Write $\crossing_G^j(H^2)$ for the number of crossing edges whose owner is at distance $j$ from its leader.
For such an edge $uv$ with owner $u$, replace either choice at $v$ in the lower component by $uv$. The inserted edge becomes a bridge, and $v$ is now at depth $j+1$ from the higher leader. We establish
\[
    2\EE[\crossing_G^j(H^2)]
    \leq n+\EE[\crossing_G^{j+1}(H^2)].
\]
The factor of two comes from the two choices that can be replaced at $v$.
To reverse a switch, fix an image and recover the previous choice at $v$.
A valid previous choice either stays inside the branch separated from the leader by the inserted bridge $uv$, or leads to another lower component.
In the first case, the branches below bridges at the same depth are disjoint and therefore sum up to $n$.
In the second case, the previous edge is a crossing edge in the image owned by $v$ at depth $j+1$, counted by $\crossing_G^{j+1}(H^2)$.
Solving the recurrence gives
\[
\EE[\crossing_G^{0}(H^2)]\leq n.
\]

\paragraph{Using one additional choice.}
It remains to handle leaders' edges to higher components.
A leader with many such edges is likely to select one of them given an additional choice, allowing us to merge its component into a higher one and remove its degree from the potential.

This motivates the augmentation $H^{2+}$. Add each $H^2$-leader's third choice only when it reaches a higher $H^2$-component.
Conditioned on $H^2$, a leader survives as a leader in $H^{2+}$ with probability equal to the fraction of its neighbors lying inside its own component or in lower components. Since each surviving leader contributes its degree to the potential,
multiplying the surviving probability by its degree gives the expected contribution to the potential, which is exactly the number of these neighbors.
Summing the internal neighbors and crossing edges to lower components gives
\[
    \EE[\Phi_G(H^{2+})] 
    \leq n+\EE[\crossing_G^0(H^2)]
    \leq 2n.
\]

\subsection{Refining the Switching Argument for \texorpdfstring{$k\geq 3$}{k >= 3}}

Since $H^{2+}\subseteq H^3$, the same potential bound holds for $H^3$. Taking $r=3$ in the first switching argument gives the desired bound for $k\geq4$.
To reach $k=3$, we must protect fewer edges, leaving room for switching. We refine the switch so that the inserted crossing edge is a bridge outside $H^{2+}$ in the image, giving
\[
    (k-2)\EE[\crossing_G(H^k)]
    \leq 3\EE[\Phi_G(H^{2+})],
    \qquad k\geq 3.
\]
The refined switching argument is given in \Cref{sec: general case k3}.

\section{Preliminary}
\label{sec: prelim}

Let $G=(V_G, E_G)$ be a simple graph, and let $N_G(v)$ denote the neighbors of $v$ in $G$.
We may assume that $G$ has no isolated vertices, as deleting them does not change the intercomponent edges and only decreases $n$.
Define the $k$-out sampling process as each node $v\in V_G$ independently and uniformly sampling $k$ edges with replacement. 
The random $k$-out subgraph $H_G^{k}$ is formed by taking the union of these sampled edges. 
These subgraphs admit a natural coupling. 
Consider the process in which each vertex independently generates an infinite sequence of independent, uniformly random incident edges
\[
\sigma_v = (e_{v,1},~ e_{v,2},~ \dots)
\]
The union of the first $k$ choices at each vertex gives $H_G^k(\sigma)$, where $\sigma:=\{\sigma_v\}_{v\in V_G}$. Thus,
\[
H_G^1(\sigma)\subseteq H_G^2(\sigma) \subseteq \cdots \subseteq H_G^k(\sigma).
\]
For fixed $k$, the ordered outcomes consisting of the first $k$ choices, including repetitions, all have probability $\prod_{v} \deg_G(v)^{-k}$. Our counting arguments use these finite prefixes as outcome, and containment is defined by prefixes. Outcome $H^k(\sigma)$ contains outcome $H^j(\sigma')$ if $\sigma'_v$ restricted to the first $j$ positions is a prefix of $\sigma_v$ for all $v\in V_G$.


Let $\CC(G)$ be the set of connected components of $G$ and $\cc(G)=|\CC(G)|$.
For a spanning subgraph $H=(V_G,E_H)\subseteq G$, define its intercomponent edges by
\[
\Crossing_G(H) :=\{uv\in E_G\mid u, v\text{ belong to different connected components of } H\}.
\]
We also call these edges crossing and write $\crossing_G(H)=|\Crossing_G(H)|$. We omit the subscript $G$ when it is clear from context.

\paragraph{Comparison of the sampling models.} \label{wor-vs-wr}
Let $H^k_{\mathrm{wr}}$ and $H^k_{\mathrm{wor}}$ denote the random
$k$-out subgraphs obtained with and without replacement, respectively. 
Using the sequences $\sigma_v$ above, take the first $k$ draws for $H^k_{\mathrm{wr}}$ and the first $\min\{k,d_G(v)\}$ distinct edges for $H^k_{\mathrm{wor}}$.
All incident edges appear almost surely (with probability $1$) in the infinite sequence, and their order of first appearance is a uniformly random permutation, so both marginals have the required distributions.
Every edge among the first $k$ draws belongs to the without-replacement sample. Hence,
\[
    H^k_{\mathrm{wr}}\subseteq H^k_{\mathrm{wor}}
    \qquad\text{almost surely}.
\]
Since having more edges can only merge components and reduce crossing edges, we have
\[
    \crossing_G(H^k_{\mathrm{wor}})
    \leq \crossing_G(H^k_{\mathrm{wr}})
    \qquad\text{almost surely}.
\]
Hence, applying \Cref{thm: geq2-out crossing}, we have $\EE[\crossing_G(H^k_{\mathrm{wor}})] \leq \EE[\crossing_G(H^k_{\mathrm{wr}})] \leq O(n/k)$, for all $k\geq 2$.

\subsection{Bridge Potential to Component Potential}

We define two potentials for spanning subgraphs of $G$ that are important in our main arguments.

\paragraph{Bridge Potential}
An edge is a bridge if deleting it increases the number of connected components. Let $\Br(G)$ denote the bridges of $G$. For nested spanning subgraphs $J\subseteq H\subseteq G$, define the \emph{bridge potential} by
\[
\Psi_G(J,H):= \sum_{e=uv\in \Br(H)\setminus E(J)} \min\{\deg_G(u), \deg_G(v)\}
\]
Each bridge's weight is at most either endpoint's degree. We can therefore charge it to either endpoint's component, using that component's maximum $G$-degree.

\paragraph{Max-degree Component Potential}
For any spanning subgraph $J=(V_G, E_J)$ of $G$, define its \emph{max-degree component potential} $\Phi$ by
\[
\Phi_G(J) := \sum_{C\in \CC(J)} \max_{v\in C} \deg_G(v)
\]
which charges each component of $J$ the largest degree of its vertices, with all degrees measured in the original graph $G$. 
Merging components can only decrease $\Phi_G$. Hence, if $J\subseteq H$, $\Phi_G(H)\leq \Phi_G(J)$.
The following lemma relates the two potentials.

\begin{lemma}
    \label{lem: bridge to component}
    Let $G$ be a simple graph and $J \subseteq H\subseteq G$ be nested spanning subgraphs of $G$. Then we have the following,
    \[
    \Psi_G(J,H) \leq \Phi_G(J)
    \]
\end{lemma}

\begin{proof}
    Each edge $e$ in $\Br(H)\setminus E(J)$ joins different connected components of $J$. Otherwise, deleting $e$ will not create new components in $H$, contradicting the definition of a bridge.
    Contract all the edges in $J$, and the edges in $\Br(H)\setminus E(J)$ form a forest joining $\CC(J)$. Root each tree arbitrarily and charge the cost $\min\{\deg_G(u), \deg_G(v)\}$ of each edge to its child component. Since each component in $\CC(J)$ has at most one parent, the following is true.
    \begin{align*}
        \sum_{e=uv\in \Br(H)\setminus E(J)} \min\{\deg_G(u), \deg_G(v)\}
        &\leq \sum_{\substack{e=uv\in \Br(H)\setminus E(J)\\ u\text{ is in the child component } C}} \deg_G(u)\\
        &\leq \sum_{C \in \CC(J) } \max_{u\in C} \deg_G(u)\\
        &= \Phi_G(J)
    \end{align*}
\end{proof}

\section{Warm-up: Switching Lemma and the near-regular case}
\label{sec: warm-up}
In this section, we demonstrate how to reduce the problem of bounding intercomponent edges to bounding the sum of the maximum degrees of each component via a key technique we call the switching argument.
We prove the following lemma in this section.
\begin{lemma}
\label{lem: switching integer}
For any integer $k\geq 1$, and integer $0\leq r\leq k$, the following is true.
For any graph $G$,
    \[
    (k-r) \EE[\crossing_G(H^{k}_G)]\leq \EE[\Phi_G(H^{r}_G)]
    \]
\end{lemma}

\noindent
By applying the following lemma on the number of components of $2$-out subgraphs due to Ghaffari, Nowicki, and Thorup \cite{ghaffari2020faster}, we can get a simple corollary bounding the number of intercomponent edges in nearly regular graphs.

\begin{lemma}[\cite{ghaffari2020faster}]
\label{lem: connected components 2-out}
    For any $n$-node graph $G$ with minimum degree $\delta$,
    $\EE[\cc(H^{2})] = O(n/\delta) $
\end{lemma}

\begin{corollary}
    For any $n$-node graph $G$ with minimum degree $\delta$ and maximum degree $\Delta$, and any $k\geq 3$,
    \[
    \EE[x(H^k)] = O\left(\frac{\Delta}{\delta} \cdot \frac{n}{k}\right)
    \]
\end{corollary}

\begin{proof}
    Apply \Cref{lem: switching integer} with $r=2$, we have $(k-2) \EE[\crossing(H^{k})]\leq \EE[\Phi(H^{2})]$. Since $\EE[\Phi(H^{2})] = \EE\left[\sum_{C\in \CC(H^2)} \max_{v\in C} \deg_G(v)\right] \leq \Delta\cdot\EE[\cc(H^2)]$, applying \Cref{lem: connected components 2-out} gives us our conclusion.
\end{proof}

The above corollary can be extended to $k \geq 2$ by extending the switching argument to real numbers $r$, by selecting the final choice with probability $r-\lfloor r\rfloor$, and complementing it with a generalization of \Cref{lem: connected components 2-out} to random subgraphs that select the second edge with non-zero probability.
However, for simplicity of warm-up, we will not expand on the details.

Motivated by applications, we now move on to the proof of \Cref{lem: switching integer}, where we use our key technique, the switching argument, a form of careful double counting.

\begin{proof}[Proof of \Cref{lem: switching integer}]
    Consider any edge $e=(u, v)\in E_G$. Write $\mathcal{E}_e$ for the event that $e$ is crossing in $H^k$. Define $v_e$ to be the designated vertex of $e$ with smaller degree in $G$, breaking ties over a globally fixed vertex order.

    Now we define a switching process at the designated vertex $v_e$ of a crossing edge $e$ in $H^k$.
    For a fixed $0\leq r\leq k$, we consider the first $r$ samples in $H^k$ the \emph{protected positions} and the other samples in $H^k$ that are not in $H^r$ the \emph{switchable positions}. 
    The number of switchable positions is $(k-r)$.

    For a fixed $H^r$ outcome $H^r(\sigma)$, we define the following forward switch operation that maps an $H^k$ outcome containing $H^r(\sigma)$ to another $H^k$ outcome containing $H^r(\sigma)$.
    For each switchable position $f$ of $v_e$, swap the position of $e$ and $f$ in the sequence of $v_e$. 
    This will make the edge $e$ a bridge selected exactly once in the resulting graph. 
    The expected number of forward switches is $(k-r)\EE[\crossing(H^k)]$.

\begin{figure}[htbp]
\centering
\begin{tikzpicture}[proof picture,x=1cm,y=1cm]
  \node[proof heading] at (3.5,2.15) {(a) Source $\sigma$: $e$ is crossing};
  \node[proof heading] at (11.35,2.15) {(b) Image $\sigma'$: $e$ is a bridge};
  \begin{scope}
    \draw[proof region] (0,-.9) rectangle (2.65,1.65);
    \draw[proof region] (4.35,-.9) rectangle (7,1.65);
    \node[proof vertex] (v) at (2,.2) {$v_e$};
    \node[proof vertex] (w) at (.65,-.25) {$w$};
    \node[proof dot] (a) at (.85,1) {};
    \node[proof vertex] (u) at (4.95,.2) {$u$};
    \node[proof dot] (b) at (6.25,1) {};
    \node[proof dot] (c) at (6.25,-.3) {};
    \draw[proof edge] (v)--(a)--(w) (u)--(b)--(c)--(u);
    \draw[proof choice] (v)--node[below,proof tag] {$f$}(w);
    \draw[proof crossing] (v)--node[above,proof tag] {$e$}(u);
    \node[proof note] at (3.5,-1.35)
      {switch at the smaller-degree endpoint $v_e$};
  \end{scope}
  \begin{scope}[xshift=7.85cm]
    \draw[proof region] (0,-.9) rectangle (7,1.65);
    \node[proof vertex] (v2) at (2,.2) {$v_e$};
    \node[proof vertex] (w2) at (.65,-.25) {$w$};
    \node[proof dot] (a2) at (.85,1) {};
    \node[proof vertex] (u2) at (4.95,.2) {$u$};
    \node[proof dot] (b2) at (6.25,1) {};
    \node[proof dot] (c2) at (6.25,-.3) {};
    \draw[proof edge] (v2)--(a2)--(w2) (u2)--(b2)--(c2)--(u2);
    \draw[proof removed] (v2)--node[below,proof tag] {$f$}(w2);
    \draw[proof bridge] (v2)--node[above,proof tag] {$e$}(u2);
    \node[proof note] at (3.5,-1.35)
      {one selecting position; $e\notin H^r$};
  \end{scope}
\end{tikzpicture}
\caption{Solid blue edges are protected positions. The dashed gold edge is crossing, hence not sampled initially. The green arrow is its unique selecting occurrence in the image. }
\label{fig:warmup-switching}
\end{figure}

    On the inverse side, consider a sampling outcome of $H^k(\sigma)$ and count the number of incoming switches that it can receive.
    Two conditions have to be met.
    First, the outcome needs to contain the fixed $H^r(\sigma)$ outcome, with the same protected choices in the same order.
    Second, there must exist a bridge edge $e \in \Br\left(H_G^k(\sigma)\right) \setminus E(H^r(\sigma))$, selected only by its designated $v_e$, and selected exactly once.

    For a sequence outcome $\sigma$, define $B_k^r(\sigma)\subseteq \Br\left(H_G^k(\sigma)\right) \setminus E(H_G^r(\sigma))$ as the set of valid bridges satisfying the above conditions. The unique occurrence of $e$ determines the position to restore. There are at most $\deg_G(v_e)$ possible previous choices that can make $e\in B_k^r(\sigma)$ a crossing edge, which is the required condition in the pre-image. Since the ordered $k$-choice sequences before and after switching have the same probability, we have,
    \[
    (k-r)\EE_{\sigma}\left[x(H^k)\right]
            \leq \EE_{\sigma }\left[\sum_{e \in B_k^r(\sigma) } \deg_G\big(v_e\big)\right]
            \leq \EE_{\sigma}[\Phi(H^r)]
    \]
    The last inequality follows from \Cref{lem: bridge to component}, since $B_k^r(\sigma)\subseteq \Br\left(H_G^k(\sigma)\right) \setminus E(H_G^r(\sigma))$ by definition, and $\deg_G(v_e) = \min_{u\in e}\deg_G(u)$ by the choice of the designated switching vertex.
\end{proof}

\section{Bounding the Component Potential and the Consequence for \texorpdfstring{$k\geq 4$}{k >= 4}}
\label{sec: general case k4}
In \Cref{sec: warm-up}, we show how the switching argument relates the number of crossing edges $\crossing_G(H^k)$ to the \emph{max-degree component} potential $\Phi_G(H^r)$. In this section, we prove the $k$-out conjecture for every $k\ge 4$ by providing a linear bound on $\EE[\Phi(H^{3})]$ improving on the crude bound $\Delta\cdot\cc(H)$.

Fix an order on the vertices by increasing degrees in $G$, breaking ties arbitrarily. We call the largest vertex of a component, which is also the one with the largest original degree, its \emph{leader}. Denote $\alpha_C$ as the leader of a component $C$. Then the component potential $\Phi_G(H)$ is the sum of the degrees of the leaders of each connected component. That is
\[
\Phi_G(H):= \sum_{C\in \CC(H)} \max_{v\in C} \deg_G(v)
= \sum_{C\in \CC(H)} \deg_G(\alpha_C)
\]
We call a component \emph{higher} if its leader is higher and \emph{lower} if its leader is lower.

\paragraph{Augmented $2$-out Graph}
To show a potential bound, our idea is to utilize an additional choice at some of the vertices to merge components together, effectively removing one previous leader from the sum toward the potential. With this idea in mind, we define the augmented $2$-out graph $H^{2+}$ in the following way. First, we sample $H^2$. Then for the leader of each connected component in $H^2$, we sample a third choice. If this edge connects to another component with a higher leader, we add that edge to $H^{2+}$. The sampling coupling applies. For the same sequence $\sigma$,
\[
H_G^2(\sigma)\subseteq H_G^{2+}(\sigma)\subseteq H_G^3(\sigma).
\]

\begin{lemma}
    \label{lem: potential augmented 2-out}
    For any simple graph $G$,
    \[
    \EE\left[\Phi_G\left(H_G^{2+}\right)\right] \leq 2n
    \]
\end{lemma}
\noindent
With this lemma, we can show our desired bound for $k\geq 4$ quite simply.

\begin{corollary}
    \label{cor: k4}
    Given any simple graph $G$, and $k\geq 4$
    \[
    \EE[\crossing(H_G^k)]\leq O(n/k).
    \]
\end{corollary}

\begin{proof}
    Since $H_G^{2+}(\sigma)\subseteq H_G^3(\sigma)$, and adding more edges can only merge components and reduce potentials, $\Phi_G\left(H_G^{2+}(\sigma)\right)\geq \Phi_G\left(H_G^{3}(\sigma)\right)$. Applying \Cref{lem: switching integer} with $r=3$ and \Cref{lem: potential augmented 2-out},
    we get
    \[
    \EE[\crossing(H^{k})]
    \leq \frac{\EE[\Phi(H^{3})]}{k-3}
    \leq \frac{\EE[\Phi(H^{2+})]}{k-3}
    \leq \frac{2n}{k-3}
    \leq O\left(\frac{n}{k}\right).
    \]
\end{proof}

\subsection{\texorpdfstring{Proof of \Cref{lem: potential augmented 2-out}: Bounded Potential on Augmented $2$-out graph}{Proof of the augmented 2-out potential bound}}

The graph $H^{2+}$ is built upon $H^2$ by adding a third randomly sampled edge from a leader of $H^2$ if that edge connects to a higher component. Hence, if a leader of $H^2$ fails to merge to a higher component, it remains a leader of $H^{2+}$ and contributes to both $\Phi(H^{2+})$ and $\Phi(H^2)$. On the other hand, if a leader of $H^2$ merges with a higher component due to its third choice, it contributes $0$ to $\Phi(H^{2+})$. The leaders of $H^{2+}$ is a subset the leaders of $H^2$, so no other vertices can contribute to $\Phi(H^{2+})$.

Conditioned on the first two choices of the sequences at each vertex, we can calculate the expected contribution of the leaders in $H^2$ to $\Phi(H^{2+})$ by bounding the probability of them staying as a leader.

Suppose $\alpha_C$ is the leader for a connected component $C$ in $H^2$. Let $d_C\leq \deg_G(\alpha_C)$ denote the number of bad neighbors of $\alpha_C$ whose connection fails to merge $C$ into a higher component. This includes exactly neighbors inside $C$ and neighbors in lower components. The probability that $\alpha_C$ remains a leader in $H^{2+}$ is
\[
\Pr\left[\alpha_C \text{ is a leader in }H^{2+} \mid H^2\right] = \frac{d_C}{\deg_G(\alpha_C)}
\]
If $\alpha_C$ remains a leader, it contributes $\deg_G(\alpha_C)$ to $\Phi(H^{2+})$. Hence, its expected contribution in this case is $\deg_G(\alpha_C) \cdot \frac{d_C}{\deg_G(\alpha_C)}= d_C$.
Hence,
\[
\EE[\Phi(H^{2+})\mid H^2] = \sum_{C\in \CC(H^2)} \deg_G(\alpha_C) \cdot \Pr\left[\alpha_C \text{ is a leader in }H^{2+} \mid H^2\right]\leq  \sum_{C\in \CC(H^2)} d_C
\]

Therefore,  it is sufficient to show that the sum of the bad neighbors of the leaders in $H^2$ is small. For each leader $\alpha_C$, we decompose the bad neighbors into internal neighbors in $C$, $N_G(\alpha_C)\cap C$, and external neighbors connecting to lower components in $H^2$, denoted by $d_C^\downarrow$. Since the components are disjoint, $\sum_{C\in \CC(H^2)} |N_G(\alpha_C)\cap C|\leq \sum_{C\in \CC(H^2)} |C| \leq n$. Then, we can write
\begin{align*}
    \sum_{C\in \CC(H^2)} d_C
    &= \sum_{C\in \CC(H^2)} \left( |N_G(\alpha_C)\cap C| ~+ d_C^\downarrow\right) \\
    &= \sum_{C\in \CC(H^2)} |N_G(\alpha_C)\cap C| + \sum_{C\in \CC(H^2)} d_C^\downarrow \\
    &\leq n+ \enlc_G(H^2)
\end{align*}
where $\enlc_G(H) := \sum_{C\in \CC(H)} d_C^\downarrow$ is the sum of the leaders' external neighbors in lower components.
Bounding the expectation of $\enlc_G(H)$ allows us to bound the expected potential of $H^{2+}$.
\[
\EE[\Phi(H^{2+})]
= \EE\left[\EE[\Phi(H^{2+})\mid H^2]\right]
\leq \EE[n+ \enlc(H^2)]
= n+ \EE[\enlc(H^2)]
\tag{*}\label{reduction to leader}
\]

\subsubsection{Leaders' external neighbors in lower components}

Now we bound $\enlc(H^2)$, the total number of external neighbors in lower components for leaders in $H^2$.
We consider all crossing edges to be owned by the endpoint in the higher component. Fix a spanning subgraph $H$ of $G$, for each vertex $v$, define the depth of $v$ in $H$ as its distance to the leader of its component. For any subgraph $H$ of $G$, define $\crossing^j_G(H)$ as the number of crossing edges whose owner is at depth $j$ of its component. Note that $\crossing^0_G(H) = \enlc_G(H)$. We will show the following relationship between the expected number of crossing edges at adjacent depths via a version of switching argument.

\begin{lemma}
    \label{lem: recurrence crossing at levels}
    For any graph $G$, and any integer $j\geq 0$,
    \[
    2\EE[\crossing_G^j(H^2)]\leq n + \EE[\crossing_G^{j+1}(H^2)]
    \]
\end{lemma}

\begin{proof}
    Fix an outcome $H^2(\sigma)$.
    Define a forward switching operation on the edges counted towards $\crossing_G^j\left(H^2(\sigma)\right)$. 
    Suppose $uv$ is a crossing edge in $H^2(\sigma)$ with $u$ in the higher component. Suppose $u$ has depth $j$. 
    We perform a switching operation that replaces each of the two existing choices of $v$ (not the owner) with $uv$. 
    Hence, for each outcome $H^2(\sigma)$, there are 
    \(
    2 \crossing_G^j\left(H^2(\sigma)\right)
    \) forward switches.

    Now consider the inverse switches. 
    Fix an outcome $H^2(\sigma)$ and consider how many forward switches it can receive. 
    A few conditions have to be met. 
    First, after switching, $uv$ is a bridge selected only by $v$, and selected only once, as it was a crossing edge before the switch. 
    Secondly, $v$ is at depth $j+1$. This is because $v$ was in a lower component before switching, so the leader of $u$'s component will remain the leader in the merged component. Since $uv$ is a bridge, every path from $v$ to the leader passes through $uv$. Hence, the depth of $v$ is exactly one more than the depth of $u$.

    To count the number of possible pre-images, we need to analyze which neighbors of $v$ could be a neighbor $w$ before the switch.
    Let $C$ be the component of $H^2(\sigma)$ containing $uv$.
    We call $v$ a \emph{bridge child} and the connected component of $v$ after the deletion of $uv$, $v$'s internal component, denoted by $C_v$. 
    There are two possible cases from which the forward switches could come, illustrated in \Cref{fig:potential-recurrence}..
    \begin{enumerate}
        \item \textbf{The $G$-neighbors of $v$ in $v$'s internal component $C_v$.} There are at most $|N_G(v)\cap C_v|$ such neighbors. Since $uv$ is a bridge, all the internal components of the bridge children at depth $j+1$ are disjoint. Hence, the total contribution here is at most 
        \[
        \sum_v |N_G(v)\cap C_v|\leq \sum_v |C_v|\leq n.
        \]

        \item \textbf{The $G$-neighbors of $v$ that are in another component of $H^2(\sigma)$ lower than $C$.} Every such edge $vw$ is a crossing edge in $H^2(\sigma)$ owned by $v$ whose depth is $j+1$. Each such edge gives at most one inverse, since the bridge and its selecting position are uniquely determined by $v$. Hence, there are at most $\crossing_G^{j+1}\left(H^2(\sigma)\right)$ such neighbors.
    \end{enumerate}
    No other sources are possible.
    Firstly, $w$ cannot be from $C\setminus C_v$. Otherwise,  restoring $vw$ after deleting $uv$ would reconnect $u$ and $v$, so $uv$ would not be a crossing edge in the pre-image.
    Secondly, $w$ cannot belong to a component higher than $C$.
    Otherwise, when $uv$ is deleted and $vw$ is restored, $v$ lives in a higher component than $u$, and $v$ will become the owner of $uv$, contradicting the definition of the switching operation.
    Thus, a fixed outcome $H^2(\sigma)$ receives at most
    \[
    n+\crossing_G^{j+1}(H^2(\sigma))
    \]
    forward switches. All ordered two-choice sequences are equally likely. Summing over these sequences and dividing by the total number, we have
    \[
    2\EE[\crossing_G^j(H^2)]\leq n + \EE[\crossing_G^{j+1}(H^2)].
    \]
\end{proof}
\begin{figure}[htbp]
\centering
\begin{tikzpicture}[proof picture,x=1cm,y=1cm]
  \node[proof heading] at (3.45,3.3) {(a) Forward: switch at the lower endpoint};
  \node[proof heading] at (11.25,3.3) {(b) Image: the two inverse cases};
  \begin{scope}
    \draw[proof region] (0,-.35) rectangle (2.25,2.8);
    \draw[proof lower] (4,-.35) rectangle (6.9,2.8);
    \node[proof leader] (a) at (1.1,2.1) {$\alpha_C$};
    \node[proof vertex] (u) at (1.1,.55) {$u$};
    \node[proof vertex] (v) at (4.6,.55) {$v$};
    \node[proof vertex] (w) at (6.2,1.9) {$w$};
    \node[proof dot] (t) at (6.2,.2) {};
    \draw[proof path] (a)--node[right,proof note] {$j$}(u);
    \draw[proof edge] (v)--(t)--(w);
    \draw[proof choice] (v)--node[left,proof tag] {$vw$}(w);
    \draw[proof crossing] (u)--node[above,proof tag] {$uv$}(v);
    \node[proof note] at (1.1,-.75) {higher component};
    \node[proof note] at (5.45,-.75) {lower component};
    \node[proof note,text width=6.9cm] at (3.45,-1.45)
      {Replace either retained choice at the lower endpoint $v$.};
  \end{scope}
  \begin{scope}[xshift=7.8cm]
    \draw[proof region] (0,1) rectangle (6.9,2.8);
    \draw[proof child] (3.65,1.1) rectangle (6.8,2.7);
    \draw[proof lower] (3.65,-1) rectangle (6.8,.45);
    \node[proof leader] (a2) at (.65,2) {$\alpha_C$};
    \node[proof vertex] (u2) at (2.25,2) {$u$};
    \node[proof vertex] (v2) at (4.25,2) {$v$};
    \node[proof vertex] (w2) at (6.05,2) {$w_{\rm in}$};
    \node[proof vertex] (z2) at (6.15,-.3) {$w_{\rm out}$};
    \draw[proof path] (a2)--node[above,proof note] {$j$}(u2);
    \draw[proof bridge] (v2)--node[above,proof tag] {$uv$}(u2);
    \draw[proof path] (v2) to[bend left=30] (w2);
    \draw[proof removed] (v2)--(w2);
    \draw[proof crossing] (v2) -- (4.25,.7)
      -- node[below,proof tag] {$vw$} (6.15,.7) -- (z2);
    \node[proof note] at (5.6,1.35) {child side $C_v$};
    \node[proof note,anchor=west,text width=1.7cm] at (3.75,-.3) {lower\\component};
    \node[proof note] at (4.25,2.5) {$j+1$};
    \node[proof note,text width=6.9cm] at (3.45,-1.65)
      {Restore $vw$ either inside $C_v$, or to a lower component.
       In the second case, $v$ owns the residual edge.};
  \end{scope}
\end{tikzpicture}
\caption{Dotted blue lines are retained paths. The green arrow is the inserted choice; dashed rose and gold edges are alternative previous choices.}
\label{fig:potential-recurrence}
\end{figure}

\begin{lemma}
    \label{lem: leader external nbr in lower component}
    For any graph $G$, $
    \EE[\enlc_G(H^2)]\leq n
    $.
\end{lemma}

\begin{proof}
    Apply \Cref{lem: recurrence crossing at levels}, we have $\EE[\crossing_G^j(H^2)]\leq \frac{n}{2} + \frac{\EE[\crossing_G^{j+1}(H^2)]}{2}$ for every integer $j\geq 0$. Hence, by a geometric sum, we have
    \begin{align*}
        \EE[\enlc_G(H^2)]
    =\EE[\crossing_G^0(H^2)]
    &\leq \frac{n}{2} + \frac{\EE[\crossing_G^{1}(H^2)]}{2}\\
    &\leq \frac{n}{2} + \frac{n/2 + \EE[\crossing_G^{2}(H^2)]/2}{2}\\
    &\leq \frac{n}{2}+ \frac{n}{4} + \cdots \\
    &\leq n.
    \end{align*}
\end{proof}
\noindent
With all ingredients needed, we can finally prove \Cref{lem: potential augmented 2-out}.

\begin{proof}[Proof of \Cref{lem: potential augmented 2-out}]
    Applying \Cref{lem: leader external nbr in lower component} and \Cref{reduction to leader}, we can conclude
    \[
    \EE[\Phi(H^{2+})]
    \leq n+ \EE[\enlc(H^2)]
    \leq n + n = 2n
    \]
\end{proof}

\section{Refined Switching to the Augmented \texorpdfstring{$2$}{2}-out Graph and the \texorpdfstring{$k\geq 3$}{k >= 3} case}
\label{sec: general case k3}

In this section, we will prove a lemma analogous to \Cref{lem: switching integer}. If we can relate the crossing edge of $H^k$ to the augmented $2$-out graph $H^{2+}$, then applying \Cref{lem: potential augmented 2-out}, we can arrive at the following version of \Cref{thm: geq2-out crossing} for $k\geq 3$.

\begin{theorem}
\label{thm: 3-out crossing}
    Let $G$ be a simple graph on $n$ vertices. For every integer $k\geq 3$,
\[
    \EE\left[\crossing_G(H_G^k)\right]
    \leq \frac{6n}{k-2}=O(n/k).
\]
\end{theorem}

\begin{lemma}
    \label{lem: switching to augmented 2-out}
    For any graph $G$, and any integer $k\geq 3$,
    \[
    (k-2)\EE[\crossing(H^{k})]\leq 3 \EE[\Phi(H^{2+})].
    \]
\end{lemma}

\begin{proof}
In this proof, we follow the convention of \Cref{lem: switching integer} that the designated vertex $v_e$ of an edge $e$ is the one with smaller degree.

Similar to the strategy in \Cref{lem: switching integer}, we define some switching/rotating operation that maps one outcome of $H^k$ to another with the following property that allows us to charge the number of switches to the bridges of $H^{k}$ that are not in $H^{2+}$ using \Cref{lem: bridge to component}. The switches should be defined such that in the image $H^k(\sigma')$, $e$ is a bridge and $e$ does not belong to $H^{2+}(\sigma')$.

    Fix a sequence $\sigma$ and an outcome $H^k(\sigma)$. We define a switching operation for all crossing edges $e$ to the $3$-rd to the $k$-th position in $\sigma_{v_e}$. There are three cases for the $3$-rd position. Let $(f_1, f_2, f_3)$ be the first three choices in order at $v_e$.
    \begin{enumerate}
        \item If $v_e$ is a leader in $H^{2+}(\sigma)$, we do not perform a third-position switch.
        \item If $v_e$ is a leader in $H^2(\sigma)$ and not a leader in $H^{2+}(\sigma)$, then we perform a rotation to $\sigma'_{v_e} = (f_3, f_2, e, \dots)$. We consider rotation a special type of switch.
        \item If $v_e$ is not a leader in $H^2(\sigma)$, then we perform the normal switch, replace the third choice with $e$, and let $\sigma'_{v_e} = (f_1, f_2, e, \dots)$.
    \end{enumerate}
    We perform the normal switches to the $4$-th to the $k$-th positions. The rotation is illustrated in \Cref{fig:three-out-rotation}.
    \begin{figure}[htbp]
\centering
\begin{tikzpicture}[proof picture,x=1cm,y=1cm]
  \node[proof heading] at (3.5,3) {(a) Before: $v_e$ leads an $H^2$-component};
  \node[proof heading] at (11.35,3) {(b) After: $v_e$ is no longer a leader};
  \begin{scope}
    \draw[proof region] (0,-1.6) rectangle (3.05,2.5);
    \draw[proof child] (4.55,.65) rectangle (7,2.5);
    \draw[proof lower] (4.55,-1.6) rectangle (7,.05);
    \node[proof leader] (v) at (2.3,.55) {$v_e$};
    \node[proof vertex] (w) at (.65,1.65) {$w$};
    \node[proof dot] (t) at (.65,-.65) {};
    \node[proof leader] (a) at (5.8,1.6) {$\alpha_D$};
    \node[proof vertex] (u) at (5.8,-.75) {$u$};
    \draw[proof edge] (w)--(t);
    \draw[proof choice] (v)--node[above left,proof tag] {$f_1$}(w);
    \draw[proof edge] (v)--node[below left,proof tag] {$f_2$}(t);
    \draw[proof bridge] (v)--node[above,sloped,proof tag] {$f_3$}(a);
    \draw[proof crossing] (v)--node[below,sloped,proof tag] {$e$}(u);
    \node[proof note] at (5.8,1) {$\alpha_D>v_e$};
    \node[proof note] at (1.5,-1.25) {$H^2$-component};
    \node[proof note,text width=7cm] at (3.5,-2.15)
      {$f_3$ already joins a higher component in $H^{2+}$;
       $e$ crosses the full $H^k$-components.};
  \end{scope}
  \begin{scope}[xshift=7.85cm]
    \draw[proof region] (0,-1.6) -- (3.05,-1.6) -- (3.05,.65)
      -- (7,.65) -- (7,2.5) -- (0,2.5) -- cycle;
    \draw[proof lower] (4.55,-1.6) rectangle (7,.05);
    \node[proof vertex] (v2) at (2.3,.55) {$v_e$};
    \node[proof vertex] (w2) at (.65,1.65) {$w$};
    \node[proof dot] (t2) at (.65,-.65) {};
    \node[proof leader] (a2) at (5.8,1.6) {$\alpha_D$};
    \node[proof vertex] (u2) at (5.8,-.75) {$u$};
    \draw[proof edge] (w2)--(t2);
    \draw[proof removed] (v2)--node[above left,proof tag] {$f_1$}(w2);
    \draw[proof edge] (v2)--node[below left,proof tag] {$f_2$}(t2);
    \draw[proof bridge] (v2)--node[above,sloped,proof tag] {$f_3$: first}(a2);
    \draw[proof choice] (v2)--node[below,sloped,proof tag] {$e$: third}(u2);
    \node[proof note] at (1.55,-1.25) {new $H^2$-component};
    \node[proof note,text width=7cm] at (3.5,-2.15)
      {$f_3$ connects $v_e$ to a higher leader already in $H^2$;
       its third choice $e$ is not added to $H^{2+}$.};
  \end{scope}
\end{tikzpicture}
\caption{Shaded regions are $H^2$-components. The dashed rose edge is the old first choice; labels mark the new positions of $f_3$ and $e$.}
\label{fig:three-out-rotation}
\end{figure}

    We count the number of forward switches.
    Each crossing edge $e$ whose designated vertex $v_e$ is a leader in $H^{2+}(\sigma)$ contributes to $(k-3)$ forward switches, while all other crossing edges contribute $(k-2)$ forward switches.
    Denote the number of crossing edges $e$ whose designated vertex $v_e$ is a leader in $H^{2+}(\sigma)$ by $\OmittedCount(\sigma)$. These are the edges that are omitted from switching. The number of forward switches from outcome sequence $\sigma$ is
    \[
    (k-3) \cdot \crossing_G\left(H^k(\sigma)\right)+\left[\crossing_G\left(H^k(\sigma)\right)-\OmittedCount(\sigma)\right]
    =(k-2)\cdot \crossing_G\left(H^k(\sigma)\right)-\OmittedCount(\sigma).
    \]
    Additionally, we have the following bound for $\OmittedCount(\sigma)$,
    \[
    \OmittedCount(\sigma) \leq \sum_{v \text{ is a leader in } H^{2+}(\sigma)} \deg_G(v) = \Phi_G\left(H^{2+}(\sigma)\right).
    \]

    Now we check that in the image $H^k(\sigma')$, $e$ is a bridge and $e$ does not belong to $H^{2+}(\sigma')$.
    This is clear for the normal cases except when $v_e$ is a leader in $H^2(\sigma)$ and not a leader in $H^{2+}(\sigma)$. In that case, we perform a rotation. The third choice $f_3$ connects $v_e$ to a higher component, and therefore belongs to $H^{2+}(\sigma)$. If we just replace $f_3$ with $e$ to make $\sigma''$, we might still connect $v_e$ to a higher component, making $e$ belong to $H^{2+}(\sigma'')$. The rotation makes sure that $f_3$ is selected in the first position, so that $v_e$ is connected to a higher component before the third choice, making $v_e$ a non-leader in $H^2(\sigma')$ and hence $e$ will not be in $H^{2+}(\sigma')$.

    Hence, the switches map all $e$ to $\Br\left(H^k(\sigma')\right)\setminus H^{2+}(\sigma')$, and we can apply \Cref{lem: bridge to component}. Switches to $e$ can come from $2(\deg_G(v_e)-1)$ possibilities if $e$ is in the third position of $\sigma'_{v_e}$ and $(\deg_G(v_e)-1)$ possibilities if $e$ is in positions $4$ onward. Since each sequence $\sigma$ has equal positive probability, summing over all sequences and dividing by the number of configurations gives
    \begin{align*}
        \sum_\sigma \Bigg( (k-2) \cdot \crossing_G\left(H^k(\sigma)\right) -\OmittedCount(\sigma)\Bigg)
        &\leq ~ \sum_{\sigma}\left(\sum_{e \in \Br\left(H^k(\sigma)\right)\setminus E\left(H^{2+}(\sigma)\right)} 2(\deg_G(v_e)-1)\right)\\
        (k-2) \cdot \EE_\sigma\left[\crossing_G(H^k)\right] -\EE_\sigma[\OmittedCount]
        &\leq ~ \EE_{\sigma}\left[\sum_{e \in \Br\left(H^k(\sigma)\right)\setminus E\left(H^{2+}(\sigma)\right)} 2(\deg_G(v_e)-1)\right]\\
        (k-2) \cdot \EE_\sigma\left[\crossing_G(H^k)\right]  &\leq ~ \EE_\sigma[\OmittedCount]+  2 \EE_{\sigma}\left[\sum_{e \in \Br\left(H^k(\sigma)\right)\setminus E\left(H^{2+}(\sigma)\right)} \min_{v\in e} \deg_G(v) \right]\\
        &\leq ~ \EE\left[\Phi_G(H^{2+})\right] + 2\EE\left[\Phi_G(H^{2+})\right]\\
        &\leq ~ 3\EE\left[\Phi_G(H^{2+})\right].
    \end{align*}
\end{proof}

Now we can prove \Cref{thm: 3-out crossing}.
\begin{proof}[Proof of \Cref{thm: 3-out crossing}]
    Apply \Cref{lem: switching to augmented 2-out}, and \Cref{lem: potential augmented 2-out}, we get
    \[
    \EE[\crossing(H^{k})]\leq ~ \frac{3}{k-2} \EE[\Phi(H^{2+})]\leq ~ \frac{6n}{k-2} = O(n/k).\qedhere
    \]
\end{proof}

\newcommand{\OutGraph}[1]{{H_{G}^{#1}}}
\newcommand{\SampleGraph}{\OutGraph{2}}
\newcommand{\FirstGraph}{\OutGraph{1}}
\newcommand{\BaseGraph}{{B}}
\newcommand{\AugGraph}{{\BaseGraph^+}}

\newcommand{\MarkedSet}{S}
\newcommand{\UnmarkedSet}{U}
\newcommand{\ReservedSet}{\mathcal R}
\newcommand{\AllowedEdges}{{E^\circ}}

\newcommand{\Choices}{{\sigma}}
\newcommand{\SelectedEdge}[2]{e_{#1,#2}}
\newcommand{\ChoiceHead}[2]{s_{#2}(#1)}
\newcommand{\PartialOutcome}{{\tau}}
\newcommand{\PartialLaw}{\nu}

\newcommand{\FixedOrder}{\pi}
\newcommand{\LeaderOrder}{\prec}
\newcommand{\Leader}[1]{\alpha_{#1}}
\newcommand{\RelevantDegree}[1]{d_{#1}}
\newcommand{\VertexDepth}[2]{h_{#1}(#2)}
\newcommand{\ChildSide}[1]{C_{#1}}
\newcommand{\SurvivalNeighbors}[1]{b_{#1}}
\newcommand{\Potential}{\Phi_{\ReservedSet}}

\newcommand{\MarkedDepthCount}[1]{M_{#1}}
\newcommand{\UnmarkedDepthCount}[1]{N_{#1}}
\newcommand{\TotalDepthCount}[1]{T_{#1}}

\newcommand{\EligibleCount}{Y}

\newcommand{\ColorCount}[2]{Z_{#1}\left(#2\right)}

\section{The \texorpdfstring{$2$}{2}-out case}
\label{sec:two-out}

In this section, we extend the result to $2$ choices with replacement. We follow the same two steps as before: relate intercomponent edges to a component potential, and then bound the expected potential. For simplicity of presentation, we only show the argument for the $2$-out case. For $k\geq3$, one can either use \Cref{thm: 3-out crossing} to get the desired bound with a better constant or extend the following argument to fit general $k$. The constants below are chosen to simplify the proof.

\begin{theorem}\label{thm:two-out}
For every simple graph $G$ on $n$ vertices, its random two-out subgraph with replacement satisfies
\[
\EE\bigl[\crossing_G(\SampleGraph)\bigr]\le500n.
\]
\end{theorem}

\subsection{Technical overview}
\label{subsec:overview}

The $3$-out argument protects two choices and some third choices to make the potential small, leaving the other third choices available for switching.
With only two choices, these two uses compete for the same second choice.
Our idea is to restrict the switching argument to a constant fraction of the crossing edges in expectation.
This loses only a constant factor in the reduction, but leads to a weaker potential in which each leader contributes only its relevant degree which counts only the neighbors used in switching.

Since we only need to switch from a random subset of vertices, we can protect two choices at the other vertices.
The graph whose potential we need to bound is therefore an augmented $1$-out graph that is already close to a $2$-out graph.
We can then adapt the earlier potential argument to bound this weaker potential in expectation by $O(n)$.

Some technical details that are different from the $3$-out case are as follows. We use a random proper coloring to define the vertices and edges that can perform switchings. As the result, we are no longer working with uniform distributions on the source and image. Hence, we need to correct the expectation calculation with the ratio of the probabilities.

\subsection{Preliminary}
\label{subsec:Preliminary}

Recall that for the $2$-out graph, each vertex $v$ independently samples two incident edges
\(
\Choices_v=(\SelectedEdge{v}{1},\SelectedEdge{v}{2})
\)
uniformly at random, with replacement. Write $\ChoiceHead{v}{i}$ for the other endpoint of the $i$-th selected edge $\SelectedEdge{v}{i}$.

\paragraph{The marked set and the base graph.}
Conditional on $\FirstGraph(\sigma)$, choose a \emph{uniformly random} proper coloring with colors $0,1,\ldots,9$.
Let $\MarkedSet$ be its zero set, call its vertices marked, and put $\UnmarkedSet=V_G\setminus\MarkedSet$. 
The set $\MarkedSet$ is an independent set in $\FirstGraph(\sigma)$. 
Uniform randomness on coloring is crucial in our analysis, since this gives each vertex probability $1/10$ of being marked and provides the probability estimates needed after switching. An arbitrary proper coloring need not satisfy these estimates.
The choice of ten is arbitrary, and any constant larger than 3 would suffice. 
We only need the information about the $\MarkedSet$, so an outcome is the pair $(\Choices,\MarkedSet)$.

We consider the \emph{base graph} $\BaseGraph(\Choices,\MarkedSet)$ that keeps the first position at every vertex and the second position of every vertex in $\UnmarkedSet$. 
Thus, all second choices at $\MarkedSet$ are omitted from $\BaseGraph$. 
Among these omitted second choices, we reserve those leaving $\MarkedSet$ for possible use in the augmentation. Denote the set of vertices with reserved choices by
\[
\ReservedSet:=\{v\in\MarkedSet:\ChoiceHead{v}{2}\in\UnmarkedSet\}.
\]

\paragraph{Leaders.}
Define the relevant degrees by
\[
\RelevantDegree{v}:=\begin{cases}
|N_G(v)\cap\UnmarkedSet|,&v\in\ReservedSet,\\
0,&v\notin\ReservedSet.
\end{cases}
\]
We assign degree zero to vertices outside $\ReservedSet$, since they have no reserved choice available for augmentation.
Fix a total order $\FixedOrder$ on the vertices, chosen independently of the sample and the coloring. 
Let $\LeaderOrder$ be the order that sorts vertices by increasing $\RelevantDegree{v}$, with ties broken according to $\FixedOrder$. 
The \emph{leader} $\Leader{C}$ of a component $C$ is its largest vertex under $\LeaderOrder$, and we compare components by their leaders similar to above.

\paragraph{Augmentation.}
Building from $\BaseGraph(\Choices,\MarkedSet)$, add the reserved second choice of each leader $\Leader{C}\in\ReservedSet$ if it reaches a higher component of $\BaseGraph(\Choices,\MarkedSet)$. 
Call the resulting graph $\AugGraph(\Choices,\MarkedSet)$. 
Then
\begin{equation}\label{eq:graph-inclusions}
\FirstGraph(\Choices)\subseteq\BaseGraph(\Choices,\MarkedSet)\subseteq\AugGraph(\Choices,\MarkedSet)\subseteq\SampleGraph(\Choices).
\end{equation}

\paragraph{Component Potential.}
We define the component potential using $\RelevantDegree{v}$ instead of $\deg_G(v)$. Since $\RelevantDegree{v}>0$ exactly on $\ReservedSet$, and $\Leader{C}=\arg\max_{v\in C} \RelevantDegree{v}$, we have
\begin{equation}\label{eq:potential}
\Potential(J):=\sum_{C\in\CC(J)}\max_{v\in C}\RelevantDegree{v} \quad
=\sum_{\substack{C\in\CC(J)\\\Leader{C}\in\ReservedSet}}\RelevantDegree{\Leader{C}}.
\end{equation}

\subsection{From intercomponent edges to component potential}
\label{subsec:crossing}

We first prove the following analogue of \Cref{lem: switching to augmented 2-out} using switching.

\begin{lemma}\label{lem:2crossing-potential}
For any simple graph $G$,
\[
\frac1{14}\EE[\crossing_G(\SampleGraph)]
\leq \frac{241}{80}\EE[\Potential(\AugGraph)].
\]
\end{lemma}

\begin{proof}
Similar to \Cref{lem: switching to augmented 2-out}, we define switches that turn crossing edges into bridges outside the augmented graph.
Fix $(\Choices,\MarkedSet)$ and an edge $e=uv\in\Crossing_G(\SampleGraph)$. 
For this procedure, order the full $\SampleGraph$-components by their largest vertices under $\FixedOrder$. 
Designate $v$ to be the endpoint in the lower component, and $u$ the endpoint in the higher component. 
This designation depends on $\Choices$ but not on $\MarkedSet$. 
This will allow us to charge the image bridge to the child component which contains the switching vertex, thereby ensuring each component is charged at most once.

The edge $e$ is eligible for switching when
\begin{equation}\label{eq:eligible}
v\in\MarkedSet,\qquad \ChoiceHead{v}{2}\in\UnmarkedSet,\qquad u\in\UnmarkedSet.
\end{equation}
Thus $v\in\ReservedSet$. For an eligible edge, write $\Choices_v=(f_1,f_2)$ and apply the following switching operation.
\begin{enumerate}
\item If $v$ is a leader in $\AugGraph(\Choices, \MarkedSet)$, omit this edge from the switching count.
\item If $v$ is a leader in $\BaseGraph(\Choices, \MarkedSet)$ but not in $\AugGraph(\Choices, \MarkedSet)$, perform the rotation
\(
\Choices'_v=(f_2,e).
\)
Since $\ChoiceHead{v}{2}\in\UnmarkedSet$, $\MarkedSet$ is still an independent set in $\FirstGraph(\Choices')$.
\item If $v$ is not a leader in $\BaseGraph(\Choices, \MarkedSet)$, perform the normal switch
\(
\Choices'_v=(f_1,e).
\)
\end{enumerate}
Leave all other choices and $\MarkedSet$ unchanged. As a result, the set $\ReservedSet$, all $\RelevantDegree{v}$, and the leader order $\LeaderOrder$ are unchanged, since both the old and the new second choices at $v$ leave $\MarkedSet$.

\paragraph{A constant fraction of crossing edges is eligible switching.}
We first count eligible crossing edges. For distinct vertices $x,y$, the random zero set $\MarkedSet$ satisfies
\begin{equation}\label{eq:pair-bound}
\Pr[x\in\MarkedSet\mid\FirstGraph(\Choices)]=\frac{1}{10},
\qquad
\Pr[y\in\MarkedSet\mid x\in\MarkedSet,\FirstGraph(\Choices)]\leq \frac{1}{8}.
\end{equation}
The first assertion follows from color symmetry, as permuting the colors of a proper coloring is also a proper coloring. 
If $x,y$ are in different components of $\FirstGraph(\Choices)$, their colors are independent, so the second probability is $1/10$. 
Otherwise, $x,y$ are connected. 
Since each component of $\FirstGraph(\Choices)$ is either a tree or contains exactly one cycle, there are at most two distinct paths from $x$ to $y$. 
Consider extending a proper coloring from $x$ to $y$ along the union of the paths from $x$ to $y$. 
There are at least $10-2=8$ colors for $y$ to choose from, and at most $1$ of them is color $0$. 
Hence, $\Pr[y\in\MarkedSet\mid x\in\MarkedSet,\FirstGraph(\Choices)]\leq \frac{1}{8}
$.  



Now condition on the entire choice sequence $\Choices$ and a crossing edge $e=vu$ designated at $v$. 
Both $\ChoiceHead{v}{2}$ and $u$ differ from $v$. Applying \Cref{eq:pair-bound} and a union bound, its eligibility probability is
\begin{align*}
    \Pr[e\text{ is eligible}\mid\Choices]
&=\Pr[v\in\MarkedSet\mid\Choices]\,
  \Pr[u,w\notin\MarkedSet\mid v\in\MarkedSet,\Choices]\\
&=\frac1{10}\left(
  1-\Pr[u\in\MarkedSet\text{ or }w\in\MarkedSet
        \mid v\in\MarkedSet,\Choices]\right)\\
&\geq\frac1{10}\left(
  1-\Pr[u\in\MarkedSet\mid v\in\MarkedSet,\Choices]
   -\Pr[w\in\MarkedSet\mid v\in\MarkedSet,\Choices]\right)\\
&\geq\frac1{10}\left(1-\frac18-\frac18\right)
 =\frac3{40}\geq\frac1{14}.
\end{align*}
Let $\EligibleCount$ count eligible crossing edges.
Conditional on $\Choices$, the set of crossing edges is fixed.
By linearity of expectation,
\begin{align}
    \EE[\EligibleCount\mid\Choices]
&=\sum_{e\in\Crossing_G(\SampleGraph)}
  \Pr[e\text{ is eligible}\mid\Choices]
\geq\frac1{14}\crossing_G(\SampleGraph),\notag
\\
\EE[\EligibleCount]
&=\EE\!\left[\EE[\EligibleCount\mid\Choices]\right]
 \geq\frac1{14}\EE[\crossing_G(\SampleGraph)]
\label{eq:eligible-count}
\end{align}
Let $\OmittedCount$ count eligible edges omitted because their
designated endpoint is an $\AugGraph$-leader.
For each component $C$ of $\AugGraph$, any such edge designated at
its leader $\Leader{C}$ goes to a neighbor in $\UnmarkedSet$.
There are at most $\RelevantDegree{\Leader{C}}$ such edges.
Summing these leader bounds over components gives
\begin{equation}\label{eq:omit-count}
\OmittedCount
\leq\sum_{C\in\CC(\AugGraph)}
       \RelevantDegree{\Leader{C}}
 =\Potential(\AugGraph).
\end{equation}
The number of performed forward switches is
$\EligibleCount-\OmittedCount$.

\paragraph{Number of inverse switches.}
Fix an image $(\Choices',\MarkedSet)$ and an edge $e$ that can receive a switch. 
Since $e$ must be a bridge in $\SampleGraph(\Choices')$, we call $e$ a \emph{receiving bridge}. 
Each $e$ must be selected only once in $\Choices'$ by one of its endpoints $v_e$, called the \emph{switching endpoint}. 
Additionally, the switching endpoint $v_e$ that selects $e$ must select $e$ at the second position. 
If $\Choices'_{v_e}=(f,e)$, the possible previous rows have the forms
\[
(f,g)\quad\hbox{for a normal switch},
\qquad
(g,f)\quad\hbox{for a rotation}.
\]
In either case, the other endpoint of $g$ must lie in $\UnmarkedSet$, so there are at most $\RelevantDegree{v_e}$ candidates for each type. 
Repeated choices are allowed, so $f=g$ is possible. 
To use these counts in the expectation, we must account for the probabilities of the outcomes.

\paragraph{Correcting contributions to the expectation for inverse switches.}
The choice sequences $\Choices$ are equally likely, but the pairs $(\Choices,\MarkedSet)$ need not be, as the conditional probability of obtaining the same zero set $\MarkedSet$ may be different for $\FirstGraph$ under different sequences $\Choices$. 
Suppose when we count forward switches from $(\Choices, \MarkedSet)$ to $(\Choices', \MarkedSet)$. Each switch contributes $\Pr[(\Choices, \MarkedSet)]$ to the expected number of forward switches. Thus, an inverse switching should be counted with a correction factor equal to their ratios of probability.

\begin{align*}
\EE[\text{\#forward switching}]=
\EE[\EligibleCount-\OmittedCount]
&=\sum_{(\Choices,\MarkedSet)}
  \Pr[(\Choices,\MarkedSet)]
  \sum_{\substack{\Choices', \, e:\\
    (\Choices,\MarkedSet,e)\mapsto(\Choices',\MarkedSet,e)}}1\\
&=\sum_{(\Choices',\MarkedSet)}
  \sum_{\substack{\Choices, \, e:\\
    (\Choices,\MarkedSet,e)\mapsto(\Choices',\MarkedSet,e)}}
  \Pr[(\Choices,\MarkedSet)]\\
&=\sum_{(\Choices',\MarkedSet)}
  \Pr[(\Choices',\MarkedSet)]
  \sum_{\substack{\Choices, \, e:\\
    (\Choices,\MarkedSet, e)\mapsto(\Choices',\MarkedSet,e)}}
  \frac{\Pr[(\Choices,\MarkedSet)]}
       {\Pr[(\Choices',\MarkedSet)]}\\
&=\sum_{(\Choices',\MarkedSet)}
  \Pr[(\Choices',\MarkedSet)]
  \sum_{\substack{\Choices, \, e:\\
    (\Choices,\MarkedSet, e)\mapsto(\Choices',\MarkedSet,e)}}
  \frac{\Pr[\MarkedSet \mid \Choices]}
       {\Pr[\MarkedSet \mid \Choices']}.
\end{align*}

We bound this correction factor now. 
Let $\ColorCount{q}{J}$ be the number of proper $q$-colorings of $J$.
For an independent set $\MarkedSet$ in $\FirstGraph$, fixing its zero class leaves nine colors on $\UnmarkedSet$, so
\begin{equation}\label{eq:zero-set-law}
\Pr[\MarkedSet\mid\FirstGraph(\Choices)]
=\frac{\ColorCount{9}{\FirstGraph(\Choices)[\UnmarkedSet]}}{\ColorCount{10}{\FirstGraph(\Choices)}}.
\end{equation}
In the case of a normal switch, $\FirstGraph(\Choices)=\FirstGraph(\Choices')$ and this probability ratio is one. 

In a rotation, the first choice at a vertex of $\MarkedSet$ is redirected to another vertex in $\UnmarkedSet$. 
This keeps the coloring proper and the induced graph $\FirstGraph[\UnmarkedSet]$ unchanged, i.e., $\FirstGraph(\Choices)[\UnmarkedSet]=\FirstGraph(\Choices')[\UnmarkedSet]$, so only the denominator in \Cref{eq:zero-set-law} changes.
Since $\FirstGraph$ can be viewed as a directed graph with exactly one directed cycle in each component, the following standard result counts its number of proper colorings. We defer the proof to \Cref{apx: coloring}.

\begin{lemma}[Coloring count for a first-choice graph]
\label{lem:first-choice-coloring-count}
Let $\ell(Q)$ denote the length of directed cycle $Q$, where two opposite directed edges form a directed cycle of length two. For every integer $q\ge 2$,
\[
\ColorCount{q}{\FirstGraph(\Choices)}
=
(q-1)^{n}
\prod_{\text{directed cycles }Q \text{ in } \FirstGraph(\Choices)}
\left(
1+\frac{(-1)^{\ell(Q)}}{(q-1)^{\ell(Q)-1}}
\right).
\]
\end{lemma}

In a rotation, changing the first choice either preserves all directed cycles or destroys the cycle it was in. It creates no new cycle. This potentially removes one term in the product in \Cref{lem:first-choice-coloring-count}. Therefore, for any rotation,
\begin{equation}\label{eq:denominator-ratio}
\frac{\Pr[\MarkedSet \mid \Choices]}{\Pr[\MarkedSet \mid \Choices']}
=\frac{\ColorCount{10}{\FirstGraph(\Choices')}}{\ColorCount{10}{\FirstGraph(\Choices)}}
\leq 1/\left(\min_{\ell} \left(
1+\frac{(-1)^{\ell}}{9^{\ell-1}}
\right)\right) \leq \frac{81}{80}.
\end{equation}

Hence, each receiving bridge $e$ contributes at most $\RelevantDegree{v_e}$ from normal switches and at most $\frac{81}{80}\RelevantDegree{v_e}$ from rotations, for a total of at most $\frac{161}{80}\RelevantDegree{v_e}$.

\paragraph{The receiving edge is a bridge of $\SampleGraph(\Choices')$ outside $\AugGraph(\Choices', \MarkedSet)$.} 

To use an argument analogous to \Cref{lem: bridge to component}, we need to show this property. We already know that each receiving edge $e$ is a bridge of $\SampleGraph(\Choices')$ as it was a crossing edge in $\SampleGraph(\Choices)$.

Recall that $\BaseGraph$ only includes second choices from $\UnmarkedSet$.
Since $e$ is selected only in the second position of
$v\notin\UnmarkedSet$ and not by any other vertices in $\Choices'$, we have $e\notin E(\BaseGraph(\Choices',\MarkedSet))$.
It remains to show that the augmentation does not add $e$. 

For a normal switch, only the second choice of $v$ changes, so $E(\BaseGraph(\Choices',\MarkedSet)) = E(\BaseGraph(\Choices,\MarkedSet))$. Since $v$ is not a $\BaseGraph(\Choices, \MarkedSet)$-leader, and the leader order did not change, $v$ is also not a $\BaseGraph(\Choices', \MarkedSet)$-leader. Hence, $e$ will also not be in $\AugGraph(\Choices', \MarkedSet)$.

For a rotation, $f_2$ reaches a higher $\BaseGraph(\Choices, \MarkedSet)$-component. Putting it at the first position connects $v$ to that higher leader even in $\BaseGraph(\Choices', \MarkedSet)$, making $v$ not a $\BaseGraph(\Choices', \MarkedSet)$-leader. Hence, the new second choice $e$ is not added to $\AugGraph(\Choices', \MarkedSet)$. Hence, $e\in\Br\left(\SampleGraph(\Choices')\right)\setminus E\left(\AugGraph(\Choices', \MarkedSet)\right)$.

\paragraph{Charging image bridges to components.} 
Contract every edge of $\SampleGraph(\Choices')$ except those in
$\Br\left(\SampleGraph(\Choices')\right)\setminus E\left(\AugGraph(\Choices', \MarkedSet)\right)$. 
The resulting graph is a forest, where each contracted vertex comes from a union of connected components of $\AugGraph(\Choices', \MarkedSet)$. 
Root each tree at the contracted vertex that contains the highest original vertex under $\FixedOrder$.
Every receiving bridge $e$ has its switching endpoint $v_e$ on the child-side of this tree, as the removal of $e$ makes the side with $u$ the side with a higher leader under $\FixedOrder$.
We can charge the contribution of each edge to the child component.
Each component of $\AugGraph(\Choices', \MarkedSet)$ receives at most one charge, since each contracted vertex receives at most one charge from its parent edge.
It follows that
\begin{align*}
\EE[\EligibleCount-\OmittedCount]
&=\sum_{(\Choices',\MarkedSet)}
  \Pr[(\Choices',\MarkedSet)]
  \sum_{\text{receiving edge } e}
  \sum_{\substack{\Choices:\\
  (\Choices,\MarkedSet,e)\mapsto(\Choices',\MarkedSet,e)}}
  \frac{\Pr[\MarkedSet\mid\Choices]}
       {\Pr[\MarkedSet\mid\Choices']}\\
&\leq 
  \sum_{(\Choices',\MarkedSet)}
  \Pr[(\Choices',\MarkedSet)]
  \sum_{\text{receiving edge } e}
   \frac{161}{80} \RelevantDegree{v_e}\\
&\leq \frac{161}{80}
  \sum_{(\Choices',\MarkedSet)}
  \Pr[(\Choices',\MarkedSet)]
  \sum_{C\in\CC\left(\AugGraph(\Choices',\MarkedSet)\right)}
  \max_{w\in C}\RelevantDegree{w}\\
&=\frac{161}{80}
  \sum_{(\Choices',\MarkedSet)}
  \Pr[(\Choices',\MarkedSet)]\Potential\left(\AugGraph(\Choices',\MarkedSet)\right)\\
&=\frac{161}{80}\EE\left[\Potential\left(\AugGraph\right)\right].
\end{align*}
Combining this with \Cref{eq:eligible-count,eq:omit-count}, we have
\begin{align*}
\frac1{14}\EE[\crossing_G(\SampleGraph)]
&\leq\EE[\EligibleCount]
=\EE[\EligibleCount-\OmittedCount]+\EE[\OmittedCount]\\
&\leq\frac{161}{80}\EE[\Potential(\AugGraph)]
+\EE[\Potential(\AugGraph)]
=\frac{241}{80}\EE[\Potential(\AugGraph)].
\end{align*}
\end{proof}

\subsection{Bounding the component potential}
\label{subsec:potential-bound}

We now bound the expected potential $\EE[\Potential(\AugGraph)]$ by examining which $\BaseGraph$-leaders survive the augmentation. 
This is analogous to \Cref{sec: general case k4}.

\begin{lemma}\label{lem:2-potential-exp}
For any simple graph $G$,
\[
\EE[\Potential(\AugGraph)]\leq 11 n .
\]
\end{lemma}

\paragraph{The expected contribution of a $\BaseGraph$-leader.} Since all leaders of $\AugGraph$ are from $\BaseGraph$, we bound the expected contribution of a $\BaseGraph$-leader.
We reveal all the randomness that is just enough to reveal $\BaseGraph$, the relevant degrees, and the leaders. 
We call these a partial outcome, denoted by 
\[
\PartialOutcome=
\Bigl(
(\SelectedEdge{v}{1})_{v\in V_G},\,
\MarkedSet,\,
\ReservedSet,\,
(\SelectedEdge{v}{2})_{v\in\UnmarkedSet}
\Bigr).
\]
These are all first choices, the sets $\MarkedSet$ and $\ReservedSet$, and the second choices at $\UnmarkedSet$. Note that $\ReservedSet$ is needed to define the relevant degrees and the leaders, and it only reveals the vertices in $\MarkedSet$ whose second choice leaves $\MarkedSet$, without revealing its actual second choice. 

Condition on $\PartialOutcome$, and consider a $\BaseGraph$-component $C$ whose leader $\Leader{C}$ belongs to $\ReservedSet$. 
Let $\SurvivalNeighbors{C}$ be the number of neighbors of $\Leader{C}$ in $\UnmarkedSet$ that lie inside $C$ or in lower $\BaseGraph$-components. Since the augmentation adds the reserved choice of $\Leader{C}$ precisely when it reaches a higher component, $\Leader{C}$ remains a leader in $\AugGraph$ with probability,
\[
\Pr[\Leader{C}\text{ is a leader in }\AugGraph\mid\PartialOutcome]
=\frac{\SurvivalNeighbors{C}}{\RelevantDegree{\Leader{C}}}.
\]
When $\Leader{C}$ remains a leader in $\AugGraph$, it contributes $\RelevantDegree{\Leader{C}}$ to the potential.
Otherwise, it contributes $0$ to the potential. Hence, the expected potential conditioned on the partial outcome $\PartialOutcome$ is
\[
\EE[\Potential(\AugGraph)\mid\PartialOutcome]
        =\sum_{\substack{C\in\CC(\BaseGraph)\\ \Leader{C}\in\ReservedSet}} \Pr[\Leader{C}\text{ is a leader in }\AugGraph
             \mid\PartialOutcome] \cdot \RelevantDegree{\Leader{C}} \quad=\sum_{\substack{C\in\CC(\BaseGraph)\\ \Leader{C}\in\ReservedSet}}\SurvivalNeighbors{C}.
\]

We split these neighbors into internal neighbors and external neighbors in lower components. The internal neighbors contribute at most $\sum_C|C|=n$. 
Let $\enlc(\BaseGraph)$ denote the total number of neighbors in $\UnmarkedSet$ that lie in lower components and are adjacent to $\BaseGraph$-leaders in $\ReservedSet$. We have
\begin{equation}\label{eq:potential-reduction}
\EE[\Potential(\AugGraph)\mid\PartialOutcome]
\leq n+\enlc(\BaseGraph),
\qquad
\EE[\Potential(\AugGraph)]
=\EE\!\left[\EE[\Potential(\AugGraph)\mid\PartialOutcome]\right]
\leq n+\EE[\enlc(\BaseGraph)].
\end{equation}
It remains to prove $\EE[\enlc(\BaseGraph)]\leq 10n$.

\subsubsection[Bounding leaders' neighbors in lower components]{Bounding leaders' neighbors in lower components with recurrence and switching}

Similar to \Cref{lem: recurrence crossing at levels}, we assign each crossing edge to its endpoint in the higher component, called its owner. We only switch edges incident to $\UnmarkedSet$. Define this set of allowed edges by
\[
\AllowedEdges:=E_G\setminus\{uv\in E_G:u,v\in\MarkedSet\}.
\]
All selected choices of $\BaseGraph$ are allowed, since $\MarkedSet$ is independent in $\FirstGraph$ and only unmarked vertices add their second choices to $\BaseGraph$.

For a vertex $u$ in a $\BaseGraph$-component $C$, let $\VertexDepth{\BaseGraph}{u}$ be its distance to $\Leader{C}$ in $\BaseGraph$. For each integer $j\geq0$, let $\MarkedDepthCount{j}(\BaseGraph)$ and $\UnmarkedDepthCount{j}(\BaseGraph)$ count allowed crossing edges whose owners are at depth $j$ and are marked and unmarked, respectively. Write
\[
\TotalDepthCount{j}(\BaseGraph)
:=\MarkedDepthCount{j}(\BaseGraph)+\UnmarkedDepthCount{j}(\BaseGraph).
\]
Every neighbor counted by $\enlc(\BaseGraph)$ corresponds to an allowed crossing edge owned by a leader in $\ReservedSet\subseteq\MarkedSet$, at depth zero. Hence,
\[
\enlc(\BaseGraph)\leq\MarkedDepthCount{0}(\BaseGraph).
\]
Note that $\MarkedDepthCount{0}$ may also include contribution fromleaders in $\MarkedSet\setminus\ReservedSet$.

We will relate these counts at adjacent depths by switching at the endpoint in the lower component, showing the following recurrences.

\begin{lemma}\label{lem:depth-recurrence}
For every integer $j\geq0$,
\begin{align}
\EE[\TotalDepthCount{j}(\BaseGraph)]
&\leq\frac87n
+\frac87\EE[\MarkedDepthCount{j+1}(\BaseGraph)]
+\frac47\EE[\UnmarkedDepthCount{j+1}(\BaseGraph)],
\label{eq:recurrence-total}\\
\EE[\MarkedDepthCount{j}(\BaseGraph)]
&\leq\frac47n
+\frac47\EE[\UnmarkedDepthCount{j+1}(\BaseGraph)].
\label{eq:recurrence-marked}
\end{align}
\end{lemma}

\begin{proof}
Fix a partial outcome $\PartialOutcome$. Suppose $e=uv$ is an allowed crossing edge whose owner $u$ has depth $j$. We define the forward switch at the lower endpoint $v$ as follows.
\begin{enumerate}
\item If $v\in\MarkedSet$, replace its first choice by $uv$, giving this switch weight $1$.
\item If $v\in\UnmarkedSet$, each of its two choices can be replaced by $uv$, giving two switches of weight $1/2$ each. 
\end{enumerate}
In the second case, since $v\in\UnmarkedSet$, changing its second choice does not change $\ReservedSet$.
Call the image $\PartialOutcome'$ with base graph $\BaseGraph'$.Since $uv$ is allowed, $\MarkedSet$ remains independent in the new first-choice graph. The relevant degrees and leader order are therefore fixed. 

We first count the forward switches with these weights. Write $p_v=1$ for $v\in\MarkedSet$ and $p_v=1/2$ for $v\in\UnmarkedSet$. Each initial edge gives total weight one, so the total forward weight from $\PartialOutcome$ is $\TotalDepthCount{j}(\BaseGraph)$.

Now consider the inverse switches. Fix an image $\PartialOutcome'$ and count how many forward switches it can receive. First, $uv$ must be a bridge of $\BaseGraph'$, selected in exactly one retained position, since it was crossing before the switch. Second, $v$ must be at depth $j+1$. This is because the old higher component is unchanged, its leader remains the leader of the merged component, and every path from $v$ to that leader passes through $uv$.
The membership of $v$ in $\MarkedSet$ or $\UnmarkedSet$ determines the type of switching it can receive. In either case, it receives switching of weight at most $1$.
Let $\ChildSide{v}$ be the component containing $v$ after deleting $uv$. The old choice $vz$ has two possible locations, illustrated in \Cref{fig:two-out-inverse-cases}.
\begin{enumerate}
\item \textbf{The $G$-neighbors of $v$ inside $\ChildSide{v}$.} Each such neighbor determines at most one inverse. The sets $\ChildSide{v}$ for bridge children like $v$ at depth $j+1$ are disjoint. Hence the total number of these inverses is at most
\[
\sum_v|N_G(v)\cap\ChildSide{v}|
\leq\sum_v|\ChildSide{v}|\leq n.
\]
\item \textbf{The $G$-neighbors of $v$ in a lower component of $\BaseGraph'$.} Since $vz$ is a crossing edge owned by $v$ at depth $j+1$. There are at most $\MarkedDepthCount{j+1}(\BaseGraph')$ such edges when $v\in \MarkedSet$ and $\UnmarkedDepthCount{j+1}(\BaseGraph')$ when $v\in \UnmarkedSet$
\end{enumerate}

\begin{figure}[htbp]
\centering
\begin{tikzpicture}[proof picture,x=1cm,y=1cm]
  \node[proof heading] at (3.35,3.65) {(a) Internal inverse: $z\in\ChildSide{v}$};
  \node[proof heading] at (10.65,3.65) {(b) Residual inverse: $z$ is outside};
  \begin{scope}
    \draw[proof region] (0,.9) rectangle (6.7,3.1);
    \draw[proof child] (3.45,1) rectangle (6.6,3);
    \node[proof leader] (a) at (.6,2.05) {$\alpha$};
    \node[proof vertex] (u) at (2.15,2.05) {$u$};
    \node[proof vertex] (v) at (4,2.05) {$v$};
    \node[proof vertex] (z) at (5.9,2.05) {$z$};
    \draw[proof path] (a)--node[above,proof tag] {$j$}(u);
    \draw[proof bridge] (v)--node[above,proof tag] {$uv$}(u);
    \draw[proof path] (v) to[bend left=35] (z);
    \draw[proof removed] (v)--(z);
    \node[proof note] at (4,2.8) {$j+1$};
    \node[proof note] at (5.1,1.3) {child side $\ChildSide{v}$};
    \node[proof note,text width=6.5cm] at (3.35,.2)
      {Restore the old choice $vz$ inside $\ChildSide{v}$.};
  \end{scope}
  \begin{scope}[xshift=7.3cm]
    \draw[proof region] (0,1.4) rectangle (6.7,3.1);
    \draw[proof child] (3.45,1.5) rectangle (6.6,3);
    \draw[proof lower] (3.45,-.75) rectangle (6.6,.6);
    \node[proof leader] (a2) at (.6,2.05) {$\alpha$};
    \node[proof vertex] (u2) at (2.15,2.05) {$u$};
    \node[proof vertex] (v2) at (4,2.05) {$v$};
    \node[proof vertex] (z2) at (5.8,-.05) {$z$};
    \draw[proof path] (a2)--node[above,proof tag] {$j$}(u2);
    \draw[proof bridge] (v2)--node[above,proof tag] {$uv$}(u2);
    \draw[proof crossing] (v2) -- (4,1.05)
      -- node[above,proof tag] {$vz$} (5.8,1.05) -- (z2);
    \node[proof note] at (4,2.8) {$j+1$};
    \node[proof note] at (5.4,2.65) {$\ChildSide{v}$};
    \node[proof note,align=left] at (4.35,-.05) {lower\\component};
    \node[proof note,text width=6.7cm] at (3.35,-1.3)
      {$v$ owns the residual edge $vz$ at depth $j+1$.};
  \end{scope}
\end{tikzpicture}
\caption{Green arrows are inserted retained choices. Dashed rose and gold edges are alternative previous choices; dotted blue lines are retained paths.}
\label{fig:two-out-inverse-cases}
\end{figure}

As in \Cref{subsec:crossing}, these outcomes are not equally likely and we need to account for the probability ratio of the partial outcomes. Let $\PartialLaw(\PartialOutcome)$ be the probability of a partial outcome. A forward switch $(\PartialOutcome,e,i)$ contributes $\PartialLaw(\PartialOutcome)p_v$. Summing these contributions by their images gives
\begin{equation}\label{eq:partial-double-count}
\EE[\TotalDepthCount{j}(\BaseGraph)]
=\sum_{\PartialOutcome'}\PartialLaw(\PartialOutcome')
\sum_{\substack{(\PartialOutcome,e,i):\\
(\PartialOutcome,e,i)\mapsto\PartialOutcome'}}
p_v\frac{\PartialLaw(\PartialOutcome)}{\PartialLaw(\PartialOutcome')}.
\end{equation}
The inner sum is over valid switches. To apply the inverse counts above, we need to bound the ratio $\PartialLaw(\PartialOutcome)/\PartialLaw(\PartialOutcome')$.

Since sampling is with replacement, the possible hidden second choices depend only on $\MarkedSet$ and $\ReservedSet$, not on the revealed choices. Since our switching above preserves both sets, so $\PartialOutcome$ and $\PartialOutcome'$ have the same number of completions to two choices. All two choice sequences have the same sampling probability, so we only need to compare the probabilities of obtaining $\MarkedSet$ from the two first-choice graphs. Writing $\FirstGraph'$ for the image first-choice graph, \Cref{eq:zero-set-law} gives
\begin{equation}\label{eq:partial-law-ratio}
\frac{\PartialLaw(\PartialOutcome)}{\PartialLaw(\PartialOutcome')}
=\frac{\ColorCount{9}{\FirstGraph[\UnmarkedSet]}}
       {\ColorCount{9}{\FirstGraph'[\UnmarkedSet]}}
 \frac{\ColorCount{10}{\FirstGraph'}}
      {\ColorCount{10}{\FirstGraph}}.
\end{equation}

If the switch changes a second position, $\FirstGraph'=\FirstGraph$ and the ratio is one. Otherwise, the new edge joins different old $\BaseGraph$-components. It therefore creates no directed cycle and destroys at most one in $\FirstGraph$.  Applying \Cref{lem:first-choice-coloring-count} as before gives
\[
\frac{\ColorCount{10}{\FirstGraph'}}{\ColorCount{10}{\FirstGraph}}
\leq\frac{81}{80}.
\]
For the nine-coloring factor, we need a lower bound on the number of colorings after the switch. If $v$ is marked, both its old and new first choices are incident to a vertex outside $\UnmarkedSet$. Neither affects the induced graph on $\UnmarkedSet$, so this factor is one.
Now suppose $v\in\UnmarkedSet$. First remove its old first choice, obtaining $\FirstGraph^-$. Every proper coloring of $\FirstGraph[\UnmarkedSet]$ is still proper in $\FirstGraph^-[\UnmarkedSet]$, since removing a choice cannot introduce a coloring constraint.
We then insert $uv$. If $u\in\MarkedSet$, this edge is absent from the induced graph on $\UnmarkedSet$, so all its colorings remain proper. 
If $u\in\UnmarkedSet$, the edge joins distinct components of $\FirstGraph^-[\UnmarkedSet]$, since $u$ and $v$ were in different $\BaseGraph$-components before the switch. The two components can be colored independently, so exactly $1/9$ of their colorings give $u$ and $v$ the same color due to color symmetry. The other $8/9$ remain proper after inserting $uv$. Thus insertion preserves at least $8/9$ of the colorings in either case, giving
\[
\ColorCount{9}{\FirstGraph'[\UnmarkedSet]}
\geq\frac89\ColorCount{9}{\FirstGraph^-[\UnmarkedSet]}
\geq\frac89\ColorCount{9}{\FirstGraph[\UnmarkedSet]}.
\]
Combining these bounds, we have
\begin{equation}\label{eq:partial-ratio}
\frac{\PartialLaw(\PartialOutcome)}{\PartialLaw(\PartialOutcome')}
\leq\frac98\cdot\frac{81}{80}
=\frac{729}{640}<\frac87.
\end{equation}
After multiplying by the switch weight $p_v$, each inverse contributes at most $8/7$ when $v$ is marked and at most $4/7$ when $v$ is unmarked. The bounds are listed in \Cref{tab:two-out-probability-correction}.
\begin{table}[htbp]
\centering
\caption{Probability ratios and position weights for the two-out potential
switching. The last column bounds the weighted contribution of one inverse.}
\label{tab:two-out-probability-correction}
\begingroup
\small
\renewcommand{\arraystretch}{1.55}
\begin{tabular}{@{}lccc@{}}
\hline
Changed position & $p_v$ & Bound on $\nu(\tau)/\nu(\tau')$
  & Bound on $p_v\nu(\tau)/\nu(\tau')$\\
\hline
$v\in\MarkedSet$, first & $1$ & $81/80$ & $81/80<8/7$\\
$v\in\UnmarkedSet$, first & $1/2$ & $729/640$ & $729/1280<4/7$\\
$v\in\UnmarkedSet$, second & $1/2$ & $1$ & $1/2<4/7$\\
\hline
\end{tabular}
\endgroup
\end{table}

We can now apply the inverse counts to upper bound the inverse contributions. There are at most $n$ internal inverses, each contributing at most $8/7$. Each residual crossing edge contributes at most $8/7$ if its owner is marked and $4/7$ if its owner is unmarked. Thus the total inverse contribution at $\PartialOutcome'$ is at most
\[
\frac87n+\frac87\MarkedDepthCount{j+1}(\BaseGraph')
+\frac47\UnmarkedDepthCount{j+1}(\BaseGraph').
\]
Substituting into \Cref{eq:partial-double-count} and taking expectations proves \Cref{eq:recurrence-total}.

For the second recurrence, count only initial edges whose owner $u$ is marked. Since the edges are allowed, their switching endpoints $v$ must be unmarked. Now every inverse has weight at most $4/7$, and every residual edge has an unmarked owner. The same argument gives
\[
\EE[\MarkedDepthCount{j}(\BaseGraph)]
\leq\frac47n+\frac47\EE[\UnmarkedDepthCount{j+1}(\BaseGraph)],
\]
which proves \Cref{eq:recurrence-marked}.
\end{proof}

We now apply the recurrences to bound the depth counts. For every $j\geq0$, we will show
\begin{equation}\label{eq:depth-bounds}
\EE[\TotalDepthCount{j}(\BaseGraph)]\leq16n,
\qquad
\EE[\MarkedDepthCount{j}(\BaseGraph)]\leq10n.
\end{equation}
We use backward induction on $j$. All depth counts are zero for $j\geq n$, which gives the base case. For the induction step, use $\TotalDepthCount{j+1}=\MarkedDepthCount{j+1}+\UnmarkedDepthCount{j+1}$ to get
\begin{align*}
\EE[\TotalDepthCount{j}]
&\leq\frac87n+\frac47\left(\EE[\TotalDepthCount{j+1}]
+\EE[\MarkedDepthCount{j+1}]\right)
\leq\frac87n+\frac47(16n+10n)=16n,\\
\EE[\MarkedDepthCount{j}]
&\leq\frac47n+\frac47\EE[\TotalDepthCount{j+1}]
\leq\frac47(n+16n)=\frac{68}{7}n\leq10n.
\end{align*}
Thus the bounds at depth $j+1$ imply those at depth $j$, proving \Cref{eq:depth-bounds}. In particular,
\[
\EE[\enlc(\BaseGraph)]
\leq\EE[\MarkedDepthCount{0}(\BaseGraph)]\leq10n.
\]

\begin{proof}[Proof of \Cref{lem:2-potential-exp}]
Applying \Cref{eq:potential-reduction} and the preceding bound gives
\[
\EE[\Potential(\AugGraph)]
\leq n+\EE[\enlc(\BaseGraph)]
\leq n+10n=11n.
\qedhere
\]
\end{proof}

We can now prove \Cref{thm:two-out}.
\begin{proof}[Proof of \Cref{thm:two-out}]
Combining \Cref{lem:2crossing-potential,lem:2-potential-exp}, we obtain
\[
\EE[\crossing_G(\SampleGraph)]
\leq14\cdot\frac{241}{80}\EE[\Potential(\AugGraph)]
\leq14\cdot\frac{241}{80}\cdot11n
=\frac{18557}{40}n<500n.
\qedhere
\]
\end{proof}

\begin{proof}[Proof of \Cref{thm: geq2-out crossing}]
The case $k\geq3$ follows from \Cref{thm: 3-out crossing}. For $k=2$, \Cref{thm:two-out} gives $\EE[\crossing_G(H_G^2)]\leq500n=O(n/k)$.
\end{proof}

\section*{Declaration of AI assistance}
\addcontentsline{toc}{section}{Declaration of AI assistance}
OpenAI Codex was used to assist with developing and checking parts of the proofs, drafting and organizing parts of the manuscript, finding relevant references, generating and revising TikZ figures, and editing the LaTeX source. The authors are responsible for the mathematical claims, references, and final content of the paper.

\bibliographystyle{alpha}
\bibliography{ref}

\appendix
\crefalias{section}{appendix}


\section{Number of proper colorings of pseudo-forest}
\label{apx: coloring}
\begin{proof}[Proof of \Cref{lem:first-choice-coloring-count}]
Each component contains exactly one directed cycle, with trees
attached to its vertices. Consider a component with $m$ vertices
whose directed cycle has length $\ell$.

First suppose $\ell\ge 3$. The cycle has
\[
(q-1)^\ell+(-1)^\ell(q-1)
\]
proper $q$-colorings. Once its colors are fixed, color the
remaining vertices in increasing distance from the cycle.
Each vertex has exactly $q-1$ choices, since only the color of
its parent is forbidden. Thus the component has
\[
(q-1)^{m-\ell}
\bigl((q-1)^\ell+(-1)^\ell(q-1)\bigr)
=
(q-1)^m
\left(
1+\frac{(-1)^\ell}{(q-1)^{\ell-1}}
\right)
\]
proper colorings.

If $\ell=2$, the two opposite arcs give a single undirected
edge, so the underlying component is a tree. Its number of
proper colorings is
\[
q(q-1)^{m-1}
=
(q-1)^m\left(1+\frac{1}{q-1}\right),
\]
which is the same formula with $\ell=2$.

Finally, colorings of distinct components can be chosen
independently. Multiplying their counts and using that their
sizes sum to $n$ proves the identity.
\end{proof}

\end{document}